\documentclass{emulateapj}\usepackage{apjfonts}
\def\Fref#1{Fig.~\ref{fig:#1}}
\def\Tref#1{Table~\ref{tab:#1}}
\def\Sref#1{\S\,\ref{sec:#1}}
\def\Eref#1{Eq.~\ref{eq:#1}}
\begin{document}
\title{Reconstructing the Guitar: Blowing Bubbles with a Pulsar Bow Shock
  Back Flow}
\author{Marten H. van Kerkwijk and Ashleigh Ingle}
\affil{Department of Astronomy and Astrophysics, University
  of Toronto, 50 St.\ George Street, Toronto, ON M5S 3H4, Canada;
  mhvk@astro.utoronto.ca}
\slugcomment{Accepted July 11, 2008, for publication in ApJ (Letters)}

\begin{abstract}
  The Guitar Nebula is an H$\alpha$ nebula produced by the interaction
  of the relativistic wind of a very fast pulsar, PSR B2224+65, with
  the interstellar medium.  It consists of a ram-pressure confined bow
  shock near its head and a series of semi-circular bubbles
  further behind, the two largest of which form the body of the
  Guitar.  We present a scenario in which this peculiar morphology is
  due to instabilities in the back flow from the pulsar bow shock.
  From simulations, these back flows appear similar to jets and their
  kinetic energy is a large fraction of the total energy in the
  pulsar's relativistic wind.  We suggest that, like jets, these flows
  become unstable some distance down-stream, leading to rapid
  dissipation of the kinetic energy into heat, and the formation of an
  expanding bubble.  We show that in this scenario the sizes,
  velocities, and surface brightnesses of the bubbles depend mostly on
  observables, and that they match roughly what is seen for the
  Guitar.  Similar instabilities may account for features seen in
  other bow shocks.
\end{abstract}
\keywords{pulsars: individual (\object{PSR~B2224+65})
      --- ISM: individual (Guitar Nebula)
      --- ISM: bubbles
      --- ISM: jets and outflows
      --- instabilities}

\section{Introduction}
\label{sec:intro}

Many pulsars travel at high speed, and the collision between their
relativistic winds and the interstellar medium leads to the formation
of bow shocks.  These shocks are observed most readily at X-ray and
optical wavelengths: the shocked relativistic wind will emit mostly
synchrotron radiation, while the shocked interstellar medium will emit
-- if it is partially neutral -- copious H$\alpha$ emission (for a
review, see \citealt{gaens06}).

Arguably the most spectacular bow shock is the Guitar Nebula, made by
one of the fastest pulsars known, PSR~B2224+65 \citep*{cordrm93}.
This H$\alpha$ nebula has, as the name implies, a guitar-like shape,
with a bright head, a faint neck, and a body consisting of two larger
bubbles (see \Fref{guitar}).  \cite{cordrm93} suggest this morphology
might reflect variations in either the pulsar energy injection rate or
the interstellar medium density.

In this {\em Letter}, we investigate whether instead the peculiar
morphology could be due to instabilities in the jet-like flow of
pulsar effluvium away from the bow shock.  Fast back flows are a
natural consequence of bow shocks: the pulsar wind is greatly heated
at the shock, which, for the usual case where cooling is slow, leads
to a high pressure that drives a flow in the only direction available,
to the back.  From simulations (e.g., \citealt{buccadz05}), the flows
seem similar to jets, being well-collimated and fast, and seem to
carry most of the pulsar wind energy.  Jet-like flows are indeed seen
in X-ray observations, which also show that only a small fraction of
the energy is radiated (for a review, \citealt{kargp08}).

So far, the simulations have not extended far to the back, but if the
back flow is similar to a jet, one might expect it to become unstable
further downstream.  From simulations of jets (e.g.,
\citealt{bodo+98}), this instability would lead to mixing with the
ambient medium and rapid dissipation of the kinetic energy into heat.
The simulations have not followed what happens beyond this initial
mixing, but it seems plausible that the material would expand rapidly
and drive a bubble.  If so, it might initially expand faster than the
pulsar motion, and gain further energy from the jet.  With time,
however, it will slow down, and once the pulsar has moved sufficiently
far ahead, the jet will become so long that it becomes unstable before
reaching the bubble, and a new bubble will be formed.  We suggest the
body of the guitar is made up of two such bubbles, while another one
has just started to form near the head.

In \Sref{model}, we describe our model in more detail, and in
\Sref{guitar} we compare it with the properties of the Guitar Nebula,
finding qualitative agreement.  In \Sref{discussion}, we discuss
implications as well as ways in which our model could be tested.

\section{Blowing Bubbles with a Bow Shock Back Flow}
\label{sec:model}

We consider a pulsar that loses energy at a rate $\dot{E}$ in the form
of a relativistic wind and moves at velocity $v_*$ through a medium of
density $\rho_0$.  The resulting bow shock will have a stand-off
distance $r_0$ given by
\begin{equation}
\frac{\dot E}{4\pi r_0^2 c}=\rho_0 v_*^2.
\label{eq:r0}
\end{equation}

The bow shock leads to a jet-like back flow carrying kinetic energy at
a rate $f_{\dot E}\dot E$, where from simulations the efficiency
factor $f_{\dot E}$ is close to unity \citep{bucc02}.  We assume the
back flow will become unstable some distance $\ell$ behind the neutron
star, rapidly mix with shocked ambient medium and dissipate its
energy, leading to the formation of a bubble.  Assuming also that the
bubble is fed more energy for some time $t_{\rm{}inj}$, and expands
adiabatically for a total time $t_{\rm exp}$, the bubble radius will
be approximately given by the Sedov-Taylor solution,
\begin{equation}
R_{\rm b} = \eta_\gamma\left(\frac{f_{\dot E}\dot E t_{\rm inj}}
                                {f_\rho \rho_0}\right)^{1/5} t_{\rm exp}^{2/5},
\label{eq:st}
\end{equation}
where $\eta_\gamma$ is a dimensionless constant of order unity that
depends on the adiabatic index of the interstellar medium and the
extent to which energy injection is instantaneous (see below), and
$f_\rho$ takes account of possible variations in density between the
head of the bow shock and the location of the bubble (for our model,
by assumption, $f_\rho\simeq1$).  

\Eref{r0} and~\ref{eq:st} both depend on the ratio $\dot E/\rho_0$,
suggesting the bubble radii can be expressed in terms of bow-shock
properties and other observables.  For this purpose, we rewrite
\Eref{r0} as,
\begin{equation}
\frac{\dot E}{\rho_0}=\frac{4\pi c}{v_*f_i^2\sin^3i}d^5\mu_*^3\theta_0^2,
\label{eq:edotbyrho}
\end{equation}
where $i$ is the inclination, $d$ the distance, $\mu_*=v_*\sin i/d$ the
proper motion, and $\theta_0=f_ir_0/d$ the angular stand-off distance
($f_i$ is a function of the inclination, with $f_{90^\circ}=1$, but
$f_i\neq\sin i$; see \citealt{gaenjs02}).  We also write
$t_{\rm exp}=(\alpha-\lambda)/\mu_*$, where $\alpha$ is the angular
separation between the center of the bubble and the pulsar and
$\lambda=\ell\sin i/d$ the angular size corresponding to the
instability length~$\ell$, and $t_{\rm 
inj}=(\beta-\Delta\lambda)/\mu_*$, where $\beta$ is the separation
between the center of the bubble and the next one closer to the
pulsar, and $\Delta\lambda$ takes into account that two bubbles can
have formed at slightly different distances behind the pulsar.  Note
that for the bubble closest to the pulsar, $t_{\rm inj}=t_{\rm exp}$
and one should replace $\beta-\Delta\lambda$ with $\alpha-\lambda$
below.  With this, the angular radius of a bubble, $\theta_{\rm
  b}=R_{\rm b}/d$, is given by,
\begin{equation}
\theta_{\rm b}=f_{\rm b}
    \left(\frac{4\pi c}{v_*}\right)^{1/5}
    \left(\theta_0^2(\alpha-\lambda)^2(\beta-\Delta\lambda)\right)^{1/5},
\label{eq:theta_b}
\end{equation}
where $f_{\rm b}=\eta_\gamma(f_{\dot E}/f_\rho f_i^2\sin^3i)^{1/5}$ is
of order unity.  One sees that for a bubble far behind the pulsar
(i.e., $\lambda\ll\alpha$ and $\Delta\lambda\ll\beta$), there is
little room to fiddle: the uncertainties in the efficiency factors,
geometry, and velocity may amount to a factor two, but they enter only
to low power.

With the sizes, the expected H$\alpha$ photon rates are given by,
\begin{equation}
n_{\alpha,\rm b} = \frac{f_\alpha 4\pi R_{\rm b}^2 v_{\rm b} n_{\rm H^0}}{4\pi d^2}
               = f_\alpha n_{\rm H^0,b}\theta_{\rm b}^2\mu_{\rm b}d,
\label{eq:nalpha_b}
\end{equation}
where $f_\alpha$ is the number of H$\alpha$ photons emitted per
neutral particle before that particle is ionised ($\sim\!0.05$ and
0.27 for case~A and B, resp., weakly dependent on velocity;
\citealt{chevr78}), $n_{\rm H^0}$ the neutral hydrogen number density,
$v_{\rm b}$ the expansion rate, and $\mu_{\rm b}$ the corresponding
proper motion,
\begin{equation}
\mu_{\rm b}=\frac{2}{5}\mu_*\frac{\theta_{\rm b}}{\alpha-\lambda},
\label{eq:mu_b}
\end{equation}
where the coefficient becomes $3/5$ for the bubble closest to the
pulsar.  Observationally, it is easiest to measure surface
brightnesses near the limbs of bubbles.  For measurement length scales
$\delta\ll\theta_{\rm b}$, one predicts $s_{\alpha,\rm limb}=
n_{\alpha,\rm b}(1/2\pi\delta\theta_{\rm b}) \sqrt{2\delta/\theta_{\rm
    b}}\propto\mu_{\rm_b}\theta_{\rm b}^{1/2}$.

We now discuss our assumptions and simplifications.  Our three main
premises are that the back flow is jet-like; that it becomes unstable;
and that the instability drives a bubble that is fed further energy
for some time.  The first two are supported by simulations: bow shocks
appear to give jet-like back flows with $r_{\rm jet}\simeq4r_0$
\citep{buccadz05}, and jets do seem to become unstable (with much of
the recent work focussing on how to prevent this from happening too
quickly; for a review, e.g., \citealt{hard04}).  Typically,
perturbations appear to grow on length scales $\ell_{\rm g}$ about
ten times the jet radius~$r_{\rm jet}$.  For instance, for
relativistic jets in a ten times denser medium, \cite{hard+98} found
$4\lesssim \ell_{\rm g}/r_{\rm jet}\lesssim15$.  For a bow shock, this
implies angular growth length scales $\lambda_{\rm g}$ in the range
$16\lesssim(\lambda_{\rm g}/\theta_0)(f_i/\sin i)\lesssim60$.  Of
course, instability will only occur after a few growth times, so the
angular distance $\lambda$ should be correspondingly larger.  We will
see in \Sref{guitar} that this is consistent with the Guitar Nebula.
It also validates an implicit assumption we made, that the bubbles do
not overtake the pulsar, i.e., that $\alpha>\theta_{\rm b}$ at all
times.  From \Eref{theta_b}, the minimum value of $\alpha-\theta_{\rm
  b}$ occurs at $\alpha=\lambda+\theta_0(3f_{\rm b}/5)^{5/2}(4\pi
c/v_*)^{1/2}$; this becomes negative only for $v_*\lesssim50\,f_{\rm
  b}^5(50\theta_0/\lambda)^2{\rm\,km\,s^{-1}}$, much smaller than the
velocity of PSR~B2224+65.

What is not clear yet, however, is whether jet instabilities could
lead to bubbles, and, if so, whether our simplified description is
justified.  In particular, in using the Sedov-Taylor solution, we
assume energy is injected (nearly) instantaneously at a single point
in a homogeneous medium, and that the expansion is adiabatic.  Of
these assumptions, the last is reasonable: the cooling time,
$t_{\rm{}cool}\approx 20000{\rm\,yr}\,(v_{\rm
  b}/100{\rm\,km\,s^{-1}})^3(\rho_0/10^{-25}{\rm\,g\,cm^{-3}})^{-1}$
\citep{koomck92}, is much longer than the $\sim\!300{\rm\,yr}$ it
takes PSR B2224+65 to cross the nebula (here, we scaled to the lowest
velocities and highest densities appropriate for the Guitar Nebula).
The others are less realistic: energy will be injected some time in a
larger, not necessarily spherical volume embedded in a medium which,
close to the axis, has been through the bow shock.  Injection over
some time should lead to a bubble that initially expands somewhat more
slowly.  Indeed, \cite{doku02} found that for continuous energy
injection, \Eref{st} can be used, but with a somewhat smaller value of
$\eta_\gamma$ (e.g., $\eta_{5/3}=0.929$ for $t_{\rm exp}=t_{\rm inj}$
instead of $\eta_{5/3}=1.152$ for $t_{\rm inj}\ll t_{\rm exp}$).  This
may be counteracted, however, by the initial expansion being in
pre-shocked, less dense medium.  At later times, our estimates should
depend less on these initial conditions, but rather on the extent to
which bubbles can be treated in isolation, when in the Guitar Nebula
they appear to have merged (\Fref{guitar}).  Overall, we conlude that
our heuristic model will only be good at the factor~2 level.

\begin{figure}
\centerline{\includegraphics[width=0.9\hsize]{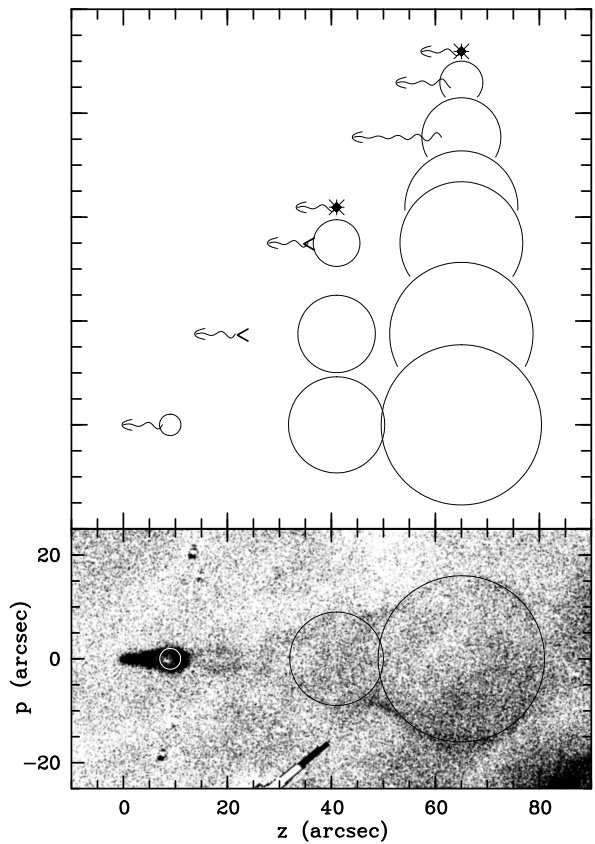}}
\caption{Reconstructing the Guitar. {\em(Top)} A possible sequence of
  events that could have led to the guitar nebula, using instabilities
  in the jet-like pulsar back flow.  The instabilities are set to
  occur at the positions marked with a star, roughly $8\arcsec$ behind
  the pulsar.  They lead to a rapid dissipation of energy and
  spherical expansion, and are fed for a while, until an instability
  occurs closer to the pulsar.  To match roughly the image of the
  guitar, we have to assume that the instabilities occurred rather
  closely together in the region marked with $<$ signs, so that no
  distinct bubbles formed. {\em(Bottom)} Continuum-subtracted
  H$\alpha$ image from the Champlane survey, with suggested locations
  of bubbles overlaid.}
\label{fig:guitar}
\end{figure}

\section{Reconstructing the Guitar}
\label{sec:guitar}

To see how well our model applies to the Guitar Nebula, we retrieved a
deep H$\alpha$ image taken on 6 Dec.\ 2000 for the Champlane survey
\citep{zhao+05}, and measured properties of what seemed the three most
obvious bubbles (see \Fref{guitar} and \Tref{guitar}): the two forming
the body of the guitar, and one just behind the pulsar (hereafter,
bottom, middle, and head).  For all bubbles, the estimates of the
angular separation from the pulsar, $\alpha$, and of the angular
radius, $\theta_{\rm b}$ are quite reliable, but for the middle
bubble, the separation to the next closest bubble, $\beta$, is
relatively poorly defined, since there may be an additional bubble in
the neck.

For our estimates, we also need the stand-off distance $\theta_0$.
\cite{chatc04} find that the shape of the bow shock -- as seen in
H$\alpha$ images taken with the {\em Hubble Space Telescope} ({\em
  HST}) -- is reproduced well with the analytic model of
\cite{wilk96}.  For data sets taken in 1994 and 2001, they infer
inclinations $i\simeq90^\circ$ and stand-off radii
$\theta_\alpha=0\farcs12\pm0\farcs04$ and $0\farcs15\pm0\farcs04$,
respectively.  Since H$\alpha$ is emitted outside the actual stand-off
distances, with $\theta_\alpha\simeq1.5\theta_0$ \citep{bucc02}, one
infers $\theta_0\simeq0\farcs09$, which we will use below.  We will
also use the observed proper motion of
$\mu_*=0\farcs182{\rm\,yr^{-1}}$ \citep{harrla93} and scale the pulsar
speed to $v_*=1500{\rm\,km\,s^{-1}}$ (where we used $i\simeq90^\circ$
and a distance $d=1.8{\rm\,kpc}$, as implied by the dispersion measure
of $35.30{\rm\,pc\,cm^{-3}}$ and the NE2001 electron density model of
\citealt{cordl02}).

\subsection{Sizes}

In matching our model to the measurements, we first note that the
existence of the head bubble implies $\lambda<9\arcsec$.  Thus, for
the bottom bubble, $\lambda\ll\alpha$ and
$\Delta\lambda\lesssim\lambda\ll\beta$.  For this case, \Eref{theta_b}
simplifies to $\theta_{\rm b}=f_{\rm b}((4\pi
c/v_*)\theta_0^2\alpha^2\beta)^{1/5}= 18\farcs3f_{\rm b}$; to match
the observed radius of $16\arcsec$ thus requires $f_{\rm b}\simeq0.9$,
close to unity as expected.

Using this value of $f_{\rm b}$ for the head bubble, we find we
require $\lambda\simeq7\farcs5$ to match the observed small size
of~$2\arcsec$.  This implies $\lambda/\theta_0\simeq80$, in line with
expectations (\Sref{model}).  It also implies the bubble formed only
recently, about $(\alpha-\lambda)/\mu_*\simeq8{\rm\,yr}$ before the
image was taken.  We return to this below.

For the middle bubble, we find that to match its observed size
requires $\beta-\Delta\lambda\simeq5\farcs5$.  This is much smaller
than for the bottom bubble, since to produce this relatively small
bubble requires much less energy.  It raises the question, though,
what happened to the energy dissipated later, outside the middle
bubble.  One possibility is that more energy was injected in the
middle bubble, but that it grew, it merged and equilibrated with the
larger bottom bubble.  If so, our above estimate of $f_{\rm b}$ would
be too large.  Clearly, we have reached the limits of applicability of
our simplistic picture of individual, spherical bubbles.

\begin{deluxetable}{lcccccc}
\tablewidth{0pt}
\tablecaption{Bubbles in the Guitar Nebula\label{tab:guitar}}
\tablehead{%
&
\colhead{$\alpha$}&
\colhead{$\theta$}&
\colhead{$\beta$}&
\colhead{$\mu$}&
&
\colhead{$\lambda$}\\
\colhead{ID}&
\colhead{$(\arcsec)$}&
\colhead{$(\arcsec)$}&
\colhead{$(\arcsec)$}&
\colhead{$(\arcsec{\rm\,yr^{-1}})$}&
\colhead{$s_{\alpha,\rm limb}$}& 
\colhead{$(\arcsec)$}}
\startdata
Head, observed\dotfill &          9  & 2    &     &      0.10 & 0.47 & \\
\phantom{Head, }model\dotfill &      & 2.1  &     &      0.15 &      & 7.5\\[1ex]
Middle, observed\dotfill &        41 & 9 &$\!\!\!<\!32$&\nodata&0.09 & \\
\phantom{Middle, }model\dotfill &    & 9.3  & 5.5 &    0.02   & 0.13 & 7.5\\[1ex]
Bottom, observed$\ldots$ &        65 & 16   & 24  &    \nodata& 0.13 & \\
\phantom{Bottom, }model\dotfill &    & 15.4 &     &    0.02   & 0.17 & 7.5
\enddata
\tablecomments{For each bubble, the first row lists measured
  parameters: $\alpha$, the angular separation between the centre of
  the bubbles and the pulsar; $\theta$, the angular radius; $\beta$,
  the separation to the next bubble (undefined for the head bubble);
  $\mu$, the expansion rate; and $s_{\alpha,\rm limb}$, the surface
  brightness near the limb (relative to that at the position of the
  pulsar, which, from the instrumental sensitivities, has a photon
  rate of $1.0\times10^{-4}{\rm\, s^{-1}\,cm^{-1}\,arcsec^{-2}}$, with
  an uncertainty of about 20\%).  The second row lists model
  assumptions and predictions, including $\lambda$, the angular
  wavelength of instability.  Empty entries indicate that observed
  values were used.  Limb surface brightnesses for the middle and
  bottom bubble were calculated relative to the observed brightness
  for the head bubble, using that $s_{\alpha,\rm
    limb}\propto\mu\theta^{1/2}$ (see \Sref{model}).}
\end{deluxetable}

\subsection{Proper motions and brightnesses}

Our model predicts expansion rates (see \Tref{guitar}).  For the head
bubble, the predicted rate is fast, $\sim\!0\farcs15{\rm\,yr^{-1}}$.
By comparing {\em HST} images, \cite{chatc04} indeed find that the
head bubble expanded between 1994 and 2001, especially to the back, at
a rate comparable if slightly slower than that predicted, of
$\sim\!0\farcs10{\rm\,yr^{-1}}$.  Interestingly, the bubble also
became brighter, consistent with the idea that it formed only
recently.  This evolution is confirmed by inspection of unpublished
{\em HST} data taken in 2006.  Furthermore, the head bubble is dimmer
in the 1992 discovery image of \cite{cordrm93} than it is in
\Fref{guitar} or in the 1995 image shown by \cite{chatc02}.

For the middle and bottom bubbles, the predicted expansion rates are
slower, $\sim\!0\farcs02{\rm\,yr^{-1}}$.  This is difficult to detect
from the ground.  It may be detectable over the 12 years spanned by
the {\em HST} images, but given the low signal-to-noise ratio, this
will require detailed modelling, which we have not attempted.

The lower proper motions for the middle and bottom bubbles also imply
predicted limb surface brightnesses about 3 times fainter than for the
head bubble.  This is roughly consistent with the observed ratio of~4
(\Tref{guitar}).

\section{Ramifications}
\label{sec:discussion}

We found that we could roughly reproduce the Guitar Nebula assuming
the jet-like back flow from the pulsar bow shock becomes unstable and
dissipates rapidly, causing expanding bubbles.  If this were to happen
generally, one might expect other sources with jets or bow shocks to
show Guitar-like bubbles, yet none appear to be known.  For jet
sources, this may not be surprising: many jets are denser than the
medium they move through, and hence more stable, and disruptions that
do occur may be difficult to distinguish from, e.g., changes in jet
orientation.  

For other bow shocks, the absence of bubbles may partly be a selection
effect: most have much larger stand-off radii than the Guitar, and
hence any bubbles would be at correspondingly larger distances, where
they might be missed, especially as they would be fainter than the bow
shock (or even invisible if the expansion velocity became too low or
if radiative effects became important; both perhaps relevant
especially for stellar wind bow shocks).  The one possible exception
is PSR B0740$-$28, which has a H$\alpha$ bow shock with a relatively
small stand-off radius of $\theta_0=1\farcs0$ as well as ``shoulders''
further behind \citep{jonesg02}.  If related to an instability, one
infers $15\lesssim\lambda/\theta_0\lesssim60$, of the same order as we
see for the Guitar.  It would be interesting to obtain deeper images
further behind the bow shock.

In some pulsar bow shocks, the shocked pulsar wind is observed
directly, by its synchrotron emission (for an overview,
\citealt{kargp08}).  For many, including the Guitar \citep{huib07},
emission is seen only close to the pulsar, likely at the pulsar wind
termination shock.  Some, however, have much longer tails.  The
longest belongs to the ``Mouse,'' associated with PSR J1747$-$2958.
This nebula, with $\theta_0\approx\!0\farcs75$, shows a bulbous
structure $\sim\!1\farcm5$ behind the pulsar (the Mouse's body), but
also a smooth, straight tail of $12\arcmin$, without a clear end
\citep{gaen+04}.  Scaling with the stand-off distances, one might
identify the Mouse's body with the equivalent of the Guitar's head
bubble.  The long tail has a size equivalent to the bottom bubble,
but, apart from changes in polarisation, shows little structure
\citep{yuseb87}.  This would seem inconsistent with any bubbles being
formed, and thus is puzzling in the context of our model.

For two other pulsar bow shocks with long tails, the observations match
expectations better.  For PSR J1509$-$5850, with
$\theta_0\approx0\farcs5$, the X-ray tail extends for
$\gtrsim\!5\farcm6$ and shows clear structure, with a change in
brightness at $1\farcm3$, a kink at~$3\arcmin$, and a bright radio
spot coincident with its end point \citep{huib07b,karg+08}.  Comparing
with the large bubbles in the Guitar, the typical length scale of
$\sim\!1\farcm5$ for the knots and kinks is about a factor 3 larger,
roughly consistent with the ratio of the stand-off distances.  For PSR
B1929+10, with $\theta_0\simeq2\farcs3$, the tail extends up to
$10\arcmin$ and again shows substantial structure, with brightenings
at $\sim\!2\arcmin$ and $\sim\!5\arcmin$, the latter coincident with a
radio feature \citep{beck+06,misapg07}.  Again scaling with the
stand-off radii, the $5\arcmin$ feature could be similar to the head
bubble in the Guitar.

Overall, we conclude that our model of instabilities in a bow shock
back flow roughly reproduces observations of the Guitar Nebula,
without the need to appeal to variations in the density of the ambient
medium, nor to energy sources beyond what is expected to be carried by
the back flow.  It also seems consistent with what is seen in other
pulsar bow shocks.  The model could be tested further both with
observations and simulations.  Observationally, one test would be to
measure the expansion velocities in the Guitar bubbles, either by
determining proper motions, or by spectroscopy (from the broad
component of the H$\alpha$ profile, as done for non-radiative shocks
in supernova remnants; \citealt{raym91}).  Given the observed
H$\alpha$ surface brightness, this would allow one to estimate the
ambient density, which should be similar to that at the location of
the bow shock in our model, but substantially lower if the bubbles
reflect density variations \citep{cordrm93,chatc04}.

Simulations of bow shocks that extend to larger scales might show
whether instabilities in fact lead to bubbles or rather to more
continuous structure, or whether perhaps the process is sufficiently
stochastic that both can occur (possibly leading to a shape like the
Guitar's neck).  If bubbles form, the simulations might also shed
light on details of the morphology, such as the closed appearance at
the back of the head and bottom bubbles.

\acknowledgements We thank Maxim Lyutikov, Ruben Krasnapolsky, and
Chris Matzner for insight in bow shock and jet instabilities, and the
referee for constructive criticism.  This research started during a
visit to the IAS, which is thanked for hospitality. It made use of ADS
and SIMBAD, and draws upon NOAO archival data from the Champlane
survey.

\bibliography{bubbles}

\end{document}